# An Intelligent Control Strategy for buck DC-DC Converter via Deep Reinforcement Learning

Chenggang Cui, Nan Yan,
Shanghai University of Electric Power
Shanghai 200082, China
E-mail：cgcui@shiep.edu.cn

Chuanlin Zhang
Shanghai University of Electric Power
Shanghai 200082, China
Email：clzhang@shiep.edu.cn

*Abstract*—As a typical switching power supply, the DC-DC converter has been widely applied in DC microgrid. Due to the variation of renewable energy generation, research and design of DC-DC converter control algorithm with outstanding dynamic characteristics has significant theoretical and practical application value. To mitigate the bus voltage stability issue in DC microgrid, an innovative intelligent control strategy for buck DC-DC converter with constant power loads (CPLs) via deep reinforcement learning algorithm is constructed for the first time. In this article, a Markov Decision Process (MDP) model and the deep Q network (DQN) algorithm are defined for DC-DC converter. A model-free based deep reinforcement learning (DRL) control strategy is appropriately designed to adjust the agent-environment interaction through the rewards/penalties mechanism towards achieving converge to nominal voltage. The agent makes approximate decisions by extracting the high-dimensional feature of complex power systems without any prior knowledge. Eventually, the simulation comparison results demonstrate that the proposed controller has stronger self-learning and self-optimization capabilities under the different scenarios.

*Keywords- DC-DC converter, constant power load(CPLs), deep Q network*

## I. INTRODUCTION

During the past few decades, the rapid development of power electronics technology has laid a splendid foundation for renewable energy sources (RESs) to be integrated into the grid[1]. However, a large amount of RESs connected to the gird will inevitably affect the stable operation of the grid due to their inherent intermittence and fluctuation. Since most distributed power sources are dc in nature (e.g., PV, fuel cells, energy storage systems, etc.), and DC loads are increasingly penetrating into microgrids[3], recent researches start to turn their eyes on DC microgrids.

For a typical DC microgrid, a significant number of power electronic converters are installed to enable high power conversion efficiency. As the most common power electronic converter, numerous research work has been conducted on DC-DC converter such as proportional-integral-derivative (PID) control[4-6], model predictive control (MPC)[7-9] and sliding mode control(SMC)[10]. But for complex system, it is difficult to design optimal controller. Therefore, intelligent strategy without accurate modeling has drawn extensive attention in recent years. In previous articles, it can be seen that the intelligent controllers including the fuzzy-PID control[11-12], fuzzy neural network control[13], brain emotional learning based intelligent controller[14] have obvious superiority in adapting themselves to the system's variable conditions.

Reinforcement Learning (RL) is an intelligent algorithm that an agent interacts with the environment by trial-and-error method to learn an optimal policy for decision-making problems[15]. As a self-learning method, RL has obtained an essential position in power system control. Research in reference[16] introduces challenges in power system control and describes reasons why the RL method can be applied for the power system. For avoiding the "high-dimension disaster", the deep reinforcement learning algorithm that combines deep learning (DL) and RL was proposed. An actor-critic framework is designed to stabilize DC-DC boost converter feeding CPLs. The actor-critic control block produces the regulatory signals[22], which are used to tune the PI controller parameters. Similarly, a deep deterministic policy gradient (DDPG) based on an ultra-local model (ULM) control strategy is discussed for voltage regulation of the DC-DC buck-boost converter[23]. In reference[24], As an auxiliary controller, the DDPG algorithm is designed to provide a compensator signal of SM observer. According to the researches and analyses mentioned above, there has been limited previous open literature to study the direct control of DC-DC converter using the DRL algorithm.

In this paper, an innovative DRL control structure is proposed to reduce voltage variation of DC-DC converter system with a wide range of operating conditions. The main contributions of this work are listed as follows:

- A DRL-based model-free intelligent controller is designed to stabilize the DC-DC buck converter with CPLs for the first time. The proposed controller achieves DC bus voltage stability without a precise mathematical model and prior knowledge.
- An MDP model is designed for the DC-DC buck converter in this paper. A novel thought is considered to design the action space, which tunes the PWM signal more effectively by adjusting a repeating sequence. Besides, the "shaping" reward procedure is adopted to hopefully guide agent towards learning an optimal policy faster[25]

## II. BACKGROUND

### A. Buck DC-DC converter

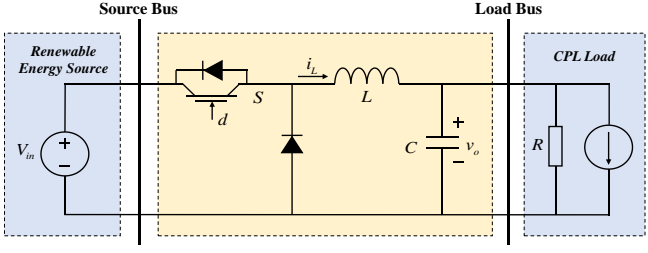

Figure 1. A Buck Converter with CPLs

The general DC-DC power buck converter is shown in Fig1. The switch $S$ works with a fixed switching frequency $f$ and a variable duty cycle $d$. The state-space model of DC-DC power buck converter is represented as equation(1).

$$\begin{cases} \dot{i}_L = \dfrac{\mu V_{in}}{L} - \dfrac{v_o}{L} \\ \dot{v}_o = \dfrac{i_L}{C} - \dfrac{v_o}{RC} - \dfrac{P_{CPL}}{Cv_o} \end{cases} \quad (1)$$

$$i_{CPL} = \dfrac{P_{CPL}}{v_o} \quad (2)$$

Where $i_L$ is inductance current, $v_o$ is output voltage, $V_{in}$ is input DC source, $\mu$ is the control signal of PWM generator, $L$ is the input inductance, $C$ is the output capacitance. $P_{CPL}$ is the constant power loads(CPLs), which provides constant power for CPLs, $i_{CPL}$ is the current generated by CPLs.

Considering the uncertainty of the system model, bus voltage variation and large CPLs changes, the state-space model of DC-DC power buck converter can be rewritten as equation(3). Where $d_1, d_2$ are the lumped uncertainties.

$$\begin{cases} \dot{i}_L = \dfrac{\mu V_{in0}}{L_0} - \dfrac{v_o}{L_0} + d_1 \\ \dot{v}_o = \dfrac{i_L}{C_0} - \dfrac{v_o}{R_0 C_0} - \dfrac{P_{CPL0}}{C_0 v_o} + d_2 \end{cases} \quad (3)$$

$$\begin{aligned} d_1 &= (\dfrac{V_{in}}{L} - \dfrac{V_{in0}}{L_0})\mu + (-\dfrac{1}{L_0} + \dfrac{1}{L}) \\ d_2 &= (\dfrac{1}{C} - \dfrac{1}{C_0})i_L + (-\dfrac{1}{CR} + \dfrac{1}{C_0 R_0})v_o - \dfrac{P_{CPL}}{Cv_o} + \dfrac{P_{CPL0}}{C_0 v_o} \end{aligned} \quad (4)$$

### B. reinforcement learning

Reinforcement learning (RL) is a self-learning intelligence method mapping environment state to action space, which obtains an optimal strategy by simultaneous trial and error. The training target is to maximize the cumulative reward(equation) received from interacting with the environment over a long term. Each time step, the agent perceives the environment state $s_t$ and makes the decision affecting the current state. The environment is then transformed into a new state $s_{t+1}$ and received a reward/punishment signal $r_t$ that estimates the state transformation. During the training process, the agent modifies the action according to the data getting from the interactions. The Markov Decision Process (MDP) satisfying the Markov property is the basic form, which means that the effects of an undertaken action in a state depend only on that state, not on the prior history.

To explore the optimal strategy, numerous algorithms are designed based on state value function $V(s)$ that quantifies the value of each state. Similarly, the state-action value function $Q^\pi(s,a)$ tells how good to perform action $a$ while in the state $s$ according to strategy $\pi$.

$$R_t = \sum_{k=0}^{\infty} \gamma^k r_{t+k+1} \quad (5)$$

Where $r$ is the reward per step, $\gamma \in (0,1)$ is the discount factor, which represents the importance of future reward to the current state.

$$V^\pi(s_t) = E_\pi\{\sum_{k=0}^{\infty} \gamma^k r_{t+k+1} \mid s_t = s\} \quad (6)$$

$$Q^\pi(s,a) = E_\pi\{\sum_{k=0}^{\infty} \gamma^k r_{t+k+1} \mid s_t = s, a_t = a\} \quad (7)$$

Generally, Q learning is applied to systems with limited actions and states. A finite Q table can be constructed to store Q value corresponding to each state-action. The agent learns to approximate the optimal control strategy by choosing the action based on Q value. The updating method of Q value is presented as equation(8).

$$Q(s,a) \leftarrow Q(s,a) + \alpha[r_{t+1} + \gamma \max_a Q(s_{t+1}, a_{t+1}) - Q(s_t, a_t)] \quad (8)$$

Where $\alpha \in (0,1)$ is the learning rate.

However, for the actual applications with the high-dimensional environment state and action, it is hard to construct an enormous Q table to store Q value. Thus, the deep reinforcement learning (DRL) combining deep learning and RL extracts the useful features from high-dimensional data. To avoid the "dimension disaster", the neural network

(NN) is adopted to capture the nonlinear relationship between the state and optimal action, which replaces the Q table to approximate the value function as an estimator.

## III. THE PROPOSED VOLTAGE CONTROL STRUCTURE FOR DC-DC CONVERTER VIA DRL

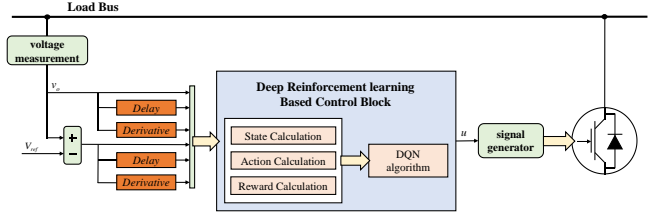

Figure 2. The Flow Chart of The Strategy Based on DRL

Considering the enormous state space and action space of the DC-DC converter system, a DRL based DC-DC converter control strategy is proposed in this paper. The deep Q network(DQN) is selected to achieve the bus voltage tracking control. The flow chart of the strategy based on DRL is shown in Fig3.

A multi-layer neural network is designed to learn the relationship between environment state and optimal action. The input is discrete state information and the output is the approximation of a state-action function $y_i$. After implementing each decision, the Q value is updated according to equation(9). The Q network is trained by minimizing the loss function changing in each interaction. And the NN parameters are updated using gradient descent method through backpropagation[26].

$$y_j = \begin{cases} r_j & \text{if episode terminate at step j+1} \\ r_j + \gamma \max_{a'} \hat{Q}(s_{j+1}, a_{j+1}; \theta^-) & \text{otherwise} \end{cases} \quad (9)$$

$$L(\theta) = E[(y_j - Q(s,a;\theta))^2] \quad (10)$$

Where $\hat{Q}(s,a,\theta)$ is the current Q value, $\theta$ is the NN parameter, $L(\theta)$ is the loss function.

The training difficulty is that small weights update will lead to large changes in the strategy. During each interaction, because the training samples do not satisfy the condition of independent and identically distribution, which has strong correlations in a short time, learning from continuous samples directly is an inefficiency method. Thus, the transitions of each step are put into an experience buffer and a mini-batch of samples of fixed size are sampled from this buffer to train the deep network. As randomly sampling from the experience buffer, the data is fully utilized in multiple times of training. This training form improves data utilization and training efficiency, and the algorithm is avoided to fall into local optimum. Furthermore, two deep networks with the same parameters are created. The first one retrieves the Q values, while the second one gives the target Q value. Periodically update the target Q network fixes the target Q value, which stabilizes the training process.

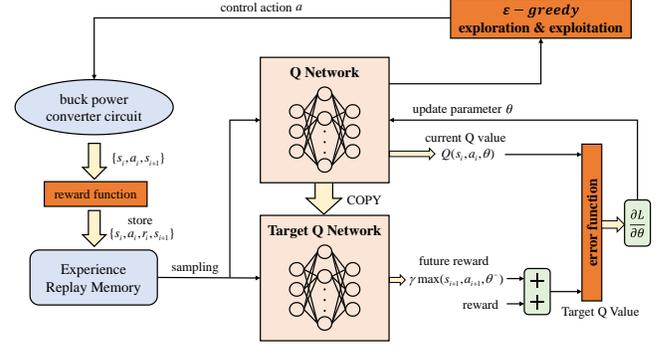

Figure 3. The Structure of DQN Algorithm

In addition, the control performance of the DRL controller largely depends on the designation of state space, action space, reward/punishment function and exploration strategy. The MDP model for the DC-DC converter control method is defined in the following part.

### A. State Space

How to design the state action is vital to any MDP solving algorithm. On the one hand, the state requires sufficient information to fully describe the system environment. The incomprehensive state can decrease the decision-making ability of the agent. Besides, the redundant state can affect the precise of extracting useful information. The control target in this paper is to stabilize the bus voltage under different operating conditions. Therefore, the output voltage $v_o(t)$ and tracking error $e(t) = v_o(t) - V_{ref}$ are considered as the underlying signals to determine the system state. Since the charging and discharging of the inductance cannot complete in an instant, the delay signal and the derivative of output voltage and tracking error are regarded as the supplementary state to characterize the system state more comprehensively. In this paper, the state is defined as follow:

$$S_t = \left\{ v_o(t), v_{o\_delay}(t), \frac{dv_o(t)}{dt}, e(t), e_{delay}(t), \frac{de(t)}{dt} \right\} \quad (11)$$

Where $v_{o\_delay}(t)$, $e_{delay}(t)$ are the delay signals of output voltage and tracking error, respectively. $\frac{dv_o(t)}{dt}$, $\frac{de(t)}{dt}$ are the derivative of output voltage and tracking error, respectively.

### B. Action Space

Lacking the established rules makes the action space design a complicated problem. The action space is designed based on the duty ratio $d \in (0,1)$ in this paper. Theoretically, continuous action space provides the different value of duty ratio $d$ for various operating conditions, which enhances the system dynamic performance. Nevertheless, continuous action space can increase the difficulty of seeking the optimal strategy, and the learning time will be longer accordingly. It is necessary to discretize the action space.

It can be learned from other mathematical model-based controller that the duty cycle $d$ is always a triangular wave with high frequency. A variable triangular wave can be obtained by tuning a fixed repeating sequence at each step. Thereby, a discrete finite action apace is constructed as $A = \{(\xi_1,\zeta_1),...,(\xi_n,\zeta_n)\}$, where $\xi$ is used for tuning the reference level of the triangular wave, $\zeta$ is used to tuning the amplitude of the triangular wave.

## C. Reward fuction

As the standard of estimating the undertaken action in the current state, the reward function achieves the communication between algorithm and control target by concretizing the control mission. The reward function represents the relationship between state and target logically guiding NN to extract the crucial information related to decision-making in the state, which influences the convergence speed and final performance of the algorithm. If a high reward is obtained by taking action $a_t$, the probability of taking action $a_t$ in state $s_t$ will increase. Otherwise, the action $a_t$ will be taken with a lower probability when the system reaches the same state $s_t$.

According to the control target, the mainline event is designed as $|e(t)|<\omega$, where $\omega$ is a constant value as small as possible. The agent can receive a positive reward when it reaches the mainline event. While for a difficult search problem, setting mainline reward can only result in sparse returns, making the algorithm hard to converge. Therefore, the auxiliary reward is added to punish the behavior of getting away from the mainline event. Regarding tracking error $e(t)$ as the potential-energy function, the gap between the current state and control target determines the potential energy. The punishment term is designed as $-\alpha|e(t)|$. The agent obtains a large punishment when the output voltage is far from the nominal voltage and the agent obtains a low punishment when the output voltage is close to the nominal voltage. Similarly, the reward term is designed as $\beta/|e(t)|$. Getting closer to the nominal voltage returns a more positive reward.

Above all, the reward function of DRL controller is designed as follow:

$$r = \begin{cases} -\alpha|e(t)| \\ \beta/|e(t)| \quad \text{if } |e(t)|<\varepsilon \end{cases} \quad (12)$$

## D. Neural Network

The NNs are designed through the state path and action path, and two paths are combined through a fully connected layer. The first hidden layer with M neurons and the second hidden layer with N neurons in the state path are designed. A hidden layer with P neurons is designed. Relu function is adopted as the activation function of each hidden layer.

## E. $\varepsilon - greedy$ strategy

Generally, 'exploiting' the action with the maximum Q value can maximize the expected reward of a certain step under any state, which can increase the convergence speed. However, due to only focusing on current optimal action the other appropriate choice beyond the experience is easily ignored by the agent. The algorithm does not reach the global optimum and fall into a local optimum. For a long term, agent explores can make better action selections in the future but it spends more time. To balance the 'exploitation' and 'exploration', the $\varepsilon - greedy$ strategy is adopted to search the action space, which is shown as equation(14).

$$a_t = \begin{cases} \arg\max_a Q(s_t,a_t) & \text{if } p<\varepsilon \\ a_{t-random} & \text{otherwise} \end{cases} \quad (13)$$

Where $p$ is a random value between 0 and 1, $\varepsilon$ is the exploration rate, $a_{t-random}$ is a random action selected from the action space.

## IV. SIMULATION AND DISCUSSION

To validate the voltage tracking ability of the DRL control strategy, the buck DC-DC converter model is established under Simulink software. The classical double-loop PI control is selected to compare with the proposed controller. The details of the buck DC-DC converter model are shown in TABLE Ⅰ. The parameters of the double-loop PI controller and DQN algorithm are shown in TABLE Ⅱ.

TABLE I. BUCK POWER CONVERTER CIRCUIT PARAMETERS

| Parameters | Definition | Value |
|---|---|---|
| $V_{in}$ | Converter input voltage | 200V |
| $V_{ref}$ | Nominal bus voltage | 100V |
| L | Nominal inductance value | 2mH |
| C | Nominal capacitance value | 150uF |
| f | Switching frequency | 20kHz |

TABLE II. CONTROLLERS PARAMETERS

| Controller | Parameters | Definition | Value |
|---|---|---|---|
| double-loop PI | $k_{cp}, k_{ci}$ | PI gains for current loop | 0.09, 35 |
| | $k_{vp}, k_{vi}$ | PI gains for voltage loop | 0.33, 40 |
| DQN | $\alpha$ | learning rate | 0.001 |
| | $\gamma$ | discount factor | 0.9 |
| | B | replay memory capacity | 1e6 |
| | b | minibatch size | 256 |
| | $\varepsilon$ | epsilon-greedy exploration | 0.1 |

## A. Test for designed parameters

For DRL algorithm, the selection of hyper-parameters has a vital impact on control performance and convergence. The learning rate, discount factor, replay memory capacity and minibatch size have been tuned after plenty times of training. The simulation results with different number of neurons and different reward parameters are analyzed in this section.

Firstly, the voltage response with different neural network is shown in Fig.5. Different number of neurons leads to different control performance. As can be seen from Fig.5, each controller with different numbers of neurons can track a certain value. The bus voltage under controller with M=8, N=16, P=16 differs by 1V from the nominal value. Only the controller with M=4, N=8, P=8 tracks the nominal value.

Then, the voltage response with different reward parameters is shown in Fig.6. Obviously, the control effect varies conspicuously under different parameters. When $\alpha=1e-3, \beta=1e-4$, the amplitude of voltage overshoot is 3.5V and the bus voltage stabilizes at 98.8V. But meanwhile, when $\alpha=1e-2, \beta=1e-3$ any obvious overshoot can be observed and the voltage reaches the desired value within 2ms.

According to the above analysis, the neurons of neural network are set as M=4, N=8, P=8. The reward parameters are set as $\alpha=1e-2, \beta=1e-3$.

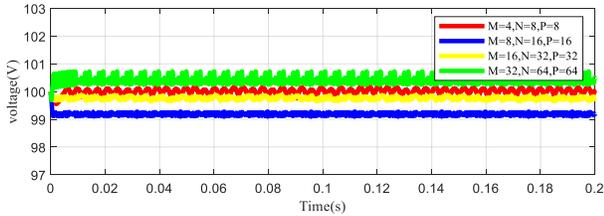

Figure 4. Voltage response with different number of neurons

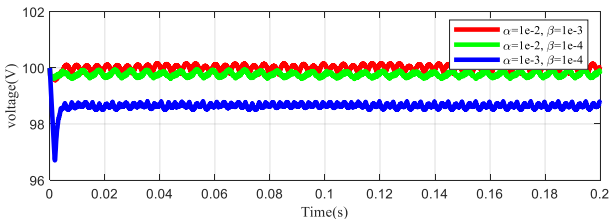

Figure 5. Voltage response with different reward parameters

## B. Test for power variation

To verify the dynamic performance of the proposed controller, the simulations with CPL variation are implemented in this section. Initially, a 300W CPL is connected. At 0.08s, the CPL increases from 300W to 500W, and at 0.14s the CPL steps back to 300W. As can be observed in Fig.7, when the disturbance occurs, the voltage can be adjusted to reach a stable state within a short time and the voltage deviation is less than 0.2% during all simulation time especially at transient time. It testifies that the proposed controller achieves the smooth transient with CPL variation.

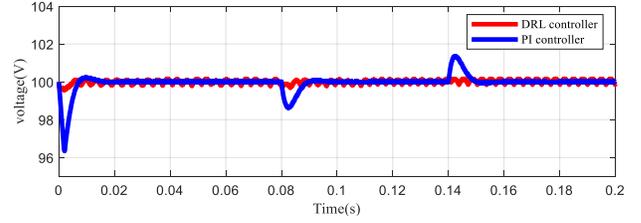

Figure 6. Voltage response with CPL variation from 300W to 500W, and from 500W to 300

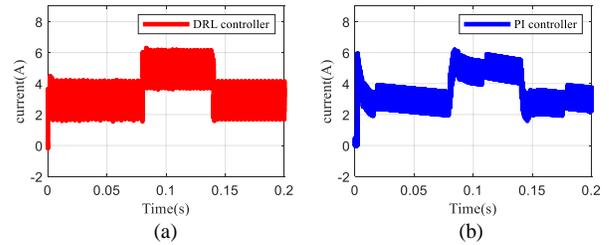

Figure 7. Current response with CPL variation from 300W to 500W, and from 500W to 300W

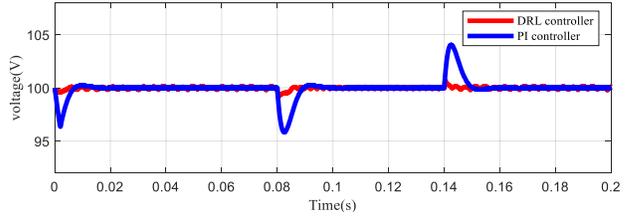

Figure 8. Voltage response with CPL variation from 300W to 900W, and from 900W to 300

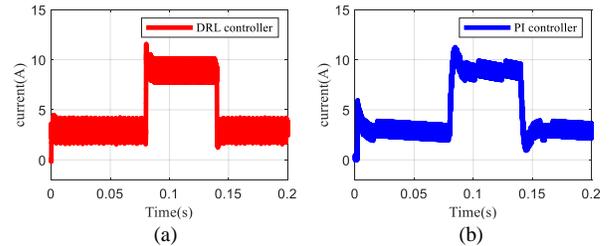

Figure 9. Current response with CPL variation from 300W to 900W, and from 900W to 300

The DRL controller is tested under the wider range of CPL variation. The CPL is initialized as 300W. At 0.08s, the CPL increases from 300W to 900W and it steps back to 300W at 0.14s. The comparison results with double-loop PI controller are presented in Fig.9. Apparently, the maximum voltage overshoot of PI controller is 4.5V while the maximum overshoot of proposed controller is 0.5V. Besides, the setting time is around 10ms of PI controller while the setting time is 5ms or less of proposed controller. Therefore, the DRL controller can achieve smooth voltage tracking with large CPL variations, which has better dynamic performance than double-loop PI controller.

## V. CONCLUSION AND FUTURE WORKS

In this paper, a DRL based controller for DC-DC buck converter feeding CPLs is proposed. The control process is described as MDP, of which the state space, action space and reward function are designed to represent the interaction process. In addition, a neural network is designed to estimate the undertaken action, which guides the agent to explore the optimal action. Then the optimal control law is learned after finite interactions. The effectiveness and the feasibility of the proposed controller are verified by numerical simulation.

Future work can focus on improving static performance of DRL control strategy. Combining DRL controller and disturbance observer can be considered. In addition, the proposed controller will be extended to a DC microgrid system. Multi-agent controller will be adopted to test the play-and-plug property.